\documentclass{aa}
\usepackage{graphicx}
\usepackage{natbib}
\usepackage{amssymb}

\def\xmm{XMM-{\it Newton}\ }
\def\sas{XMM-SAS }
\def\epic{ EPIC }
\def\clxmm{\object{2XMM J083026.2+524133} }
\def\rosat{ {\it ROSAT}\ }
\def\clrosat{ \object{CL J1226.9+3332} }

\newcommand{\Fig}[1]{Fig.\,\ref{#1}}

\def\clshort{2XMM\,0830\ }
\def\xspec{{\tt XSPEC\ }}
\def\mekal{{\tt MEKAL\ }}

\begin{document}

   \title{2XMM J083026+524133: \\ The most X-ray luminous cluster at redshift 1}

   \author{G. Lamer\inst{1}
           \and M. Hoeft\inst{1}
           \and J. Kohnert\inst{1}
           \and A. Schwope\inst{1} 
           \and J. Storm\inst{1} }

   \offprints{G. Lamer}

   \institute{Astrophysikalisches Institut Potsdam,
               An der Sternwarte 16,
               D-14482 Potsdam, Germany\\
               \email{glamer@aip.de}
                }

   \date{Received; accepted}

   \abstract
    {}
    {In the distant universe X-ray luminous clusters of galaxies are rare objects.
     Large area surveys are therefore needed to probe the high luminosity end of
     the cluster population at redshifts $z \gtrsim 1$.
    }
    {We correlated extended X-ray sources from the second XMM-Newton source catalogue (2XMM)  
     with the SDSS in order to identify new clusters of galaxies.  Distant cluster candidates in empty SDSS
     fields were imaged in the $r$- and $z$-bands with the Large Binocular Telescope. 
     We extracted the X-ray spectra of the cluster candidates and fitted thermal plasma models to the data.
     }
    {We determined the redshift $0.99\pm0.03$ for \clxmm from its X-ray spectrum. 
     With a bolometric luminosity of $1.8 \times 10^{45} {\rm erg \, s^{-1}}$ this is
     the most X-ray luminous cluster at redshifts $z \geq 1$.
     We measured a gas temperature of $8.2 \pm 0.9$ keV 
     and estimate a cluster mass $M_{500} = 5.6 \times 10^{14} M_{\odot}$.  
     The optical imaging revealed a rich cluster of galaxies.   
     }
    {}


     \authorrunning{G. Lamer et al.}
     \titlerunning{\clxmm}

     \maketitle

\section{Introduction}
\label{intro}

Clusters of galaxies play an important role as cosmological probes and
as laboratories for the evolution of galaxies.
There is observational evidence that the cosmological evolution  of the cluster 
X-ray luminosity function makes very massive, X-ray luminous clusters rare 
objects at high redshifts  \citep[e.g.][]{Henry,rosati02}. 
At the same time, simulations show that the  number counts of the most 
massive distant clusters depend strongly on the cosmological parameters, 
making them the most interesting cluster sample for cosmological studies \citep[e.g.][]{borgani,romer01}.

After the identification of the \rosat surveys only a handful of clusters at redshifts beyond
$z = 1$ were known, since only few \rosat observations were deep enough to reach
the necessary flux limits. The identification of  serendipitously detected 
\xmm sources pushed the redshift limit for X-ray selected clusters beyond $z=1.4$, where
two clusters with X-ray luminosities of several $10^{44} {\rm erg s^{-1}}$ have been discovered 
\citep{mullis,stanford}.
Clusters at redshifts $ z\sim 1$ are now routinely identified from serendipitous \xmm sources
or in dedicated survey fields \citep[e.g.][]{pierre06,cosmos}. 
However, the majority of these clusters have 
luminosities of only few times $10^{43} {\rm erg \, s^{-1}}$ or less,
placing them in the transition regime between clusters and groups of galaxies. 
In order to probe the high-luminosity part of the distant cluster
population, large area surveys are needed to find high luminosity, distant clusters.
To date the largest medium sensitivity X-ray survey is the serendipitous \xmm survey, 
which contains all sources found with the EPIC cameras in the public \xmm observations.  

In section \ref{selection} of this paper we describe the selection of cluster candidates from \xmm sources.
The analysis of the X-ray data and optical observations of a new distant and very luminous cluster is described in 
sections \ref{xray} and  \ref{lbt} respectively. Section \ref{conclusion} concludes the paper with a brief
discussion and outlook.

The cosmological parameters  $\Omega_{\rm M}=0.3$, $\Omega_{\Lambda}=0.7$ and $H_0=71\; {\rm km/(s Mpc)}$
have been used throughout this paper, unless mentioned otherwise.
With these parameters the linear scale at $z=1.0$ is $7.9 {\rm kpc/''}$. 

\section{Cluster candidate selection}
\label{selection} 

We have started a programme to identify the extended source content of the second
 \xmm catalogue \citep[2XMM][]{watson}. 
2XMM is based on a compilation of 
 $\sim 3500$ \epic observations performed before May 2007.   
It covers a sky area of $\sim 360$ square degrees and comprises 192,000 unique X-ray sources.
The source detection software, based on the \sas version 7.1, was configured to discern point like 
and extended X-ray sources and a $\beta$-model profile is fitted to the images of extended
sources. 
Since  the spurious detection rate among the $\sim 20000$ extended sources in 2XMM is relatively high,
we have cleaned the sample of extended sources for obviously spurious detections by means of visual 
inspection of the X-ray images. 

We have correlated the confirmed extended sources with the spectroscopic and photometric 
databases of the Sloan Digital Sky Survey (SDSS). The 2XMM survey area within the SDSS DR6 footprint
is $\sim 120$ square degrees.
The results on the 2XMM cluster candidates 
detected in the SDSS will be reported elsewhere.
For a relatively small number of extended X-ray sources no cluster or galaxy counterpart
is visible in the SDSS images. These sources can be considered as candidates for distant
clusters beyond  redshifts of $\sim 0.8$.
The cluster \clxmm was the brightest X-ray source among these ``empty field'' candidates.

\section{X-ray data }
\label{xray}

   \begin{figure}
     \includegraphics[width=8.8cm]{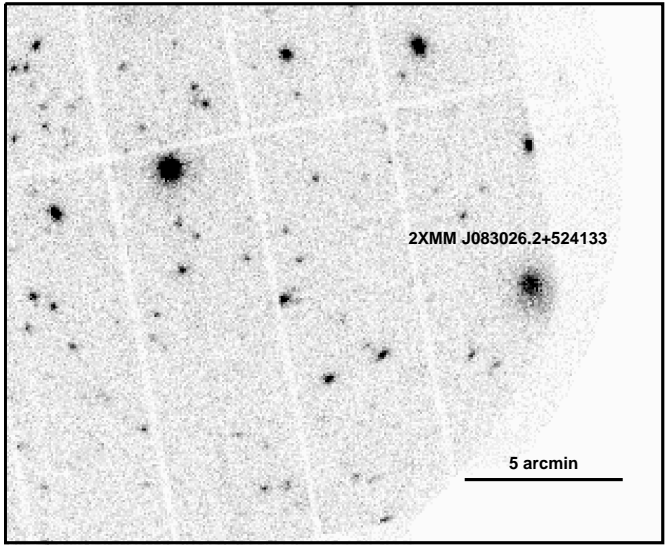}
     \caption{
       EPIC MOS1 + PN image of observation 0092800201 in the 0.2-4.5 keV energy band.
       } 
     \label{epic} 
   \end{figure}

   \clxmm has been serendipitously detected in XMM EPIC observations of
   the quasar APM 08279+5255 (Observation IDs 0092800101, 0092800201)
   and entered the 2XMM source catalogue as an extended X-ray source.
   The extended nature was clearly confirmed by our visual screening
   (see \Fig{epic}). Some catalogue parameters of the source, henceforth referred to as \clshort
   are summarised in Table~1.

   \begin{table}[b]
     \begin{tabular}{lll}
     \label{xmmtab}
     ObsID                                                       &  0092800101           &   0092800201         \\[.4ex]
     \hline
     PN paramater \\
     \rule{1.5mm}{0mm}  ONTIME [s]                               &  13896                &   72133              \\
     \rule{1.5mm}{0mm}  OFFAX  [$''$]                            &  13.2                 &   10.9               \\
     \rule{1.5mm}{0mm}  CR .2-12 keV [s$^{-1}$]                  &  $0.148 \pm .009$    & $0.115 \pm .003$    \\
     PN flux \\
     \rule{1.5mm}{0mm}   .2-12 keV [$10^{-13}{\rm erg/ cm^2 s}$]  &  $4.79 \pm 0.99 $     & $3.46 \pm 0.28 $     \\
     \rule{1.5mm}{0mm}   .5-2 keV  [$10^{-13}{\rm erg/ cm^2 s}$]   &  $1.40 \pm 0.10 $     & $1.11 \pm 0.03 $     \\
     \rule{1.5mm}{0mm}  COUNTS                                   &  $627 \pm  37$        & $3457  \pm  84$      \\ [.4ex]
     \hline
    \end{tabular}
    \caption{ 
      \clshort : 2XMM catalogue source parameters
    }
  \end{table}

   With more than 4000 detected source counts in the EPIC PN camera the
   source is bright enough to attempt a redshift determination based on
   its X-ray spectrum. 
   We extracted the source counts from the merged data files and created
   the spectra, the detector response matrix, and the effective area
   using XMM SAS version 7.1. In the MOS2 image the cluster centre
   unfortunately lies on a gap between two CCDs. Therefore we excluded
   this camera from our analysis. We varied the aperture of the source
   extraction region and found that with a radius of $45''$ the
   cluster redshift and temperature are determined with smallest
   uncertainties. 
   \\

   \begin{figure}
   \begin{center}
   \includegraphics[width=0.32\textwidth,angle=-90]{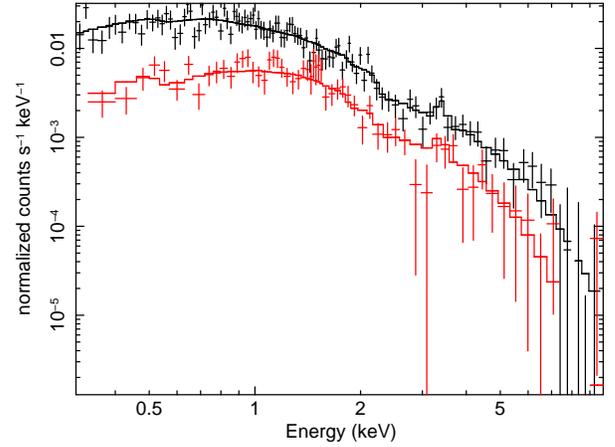}
   \end{center}
   \caption{
      EPIC PN (upper spectrum in black) and MOS1 (lower spectrum in red) data with
      best fitting \mekal model.
      } 
   \label{spectrum} 
   \end{figure}

   \begin{figure*}
   \begin{center}
   \includegraphics[width=0.33\textwidth,angle=-90]{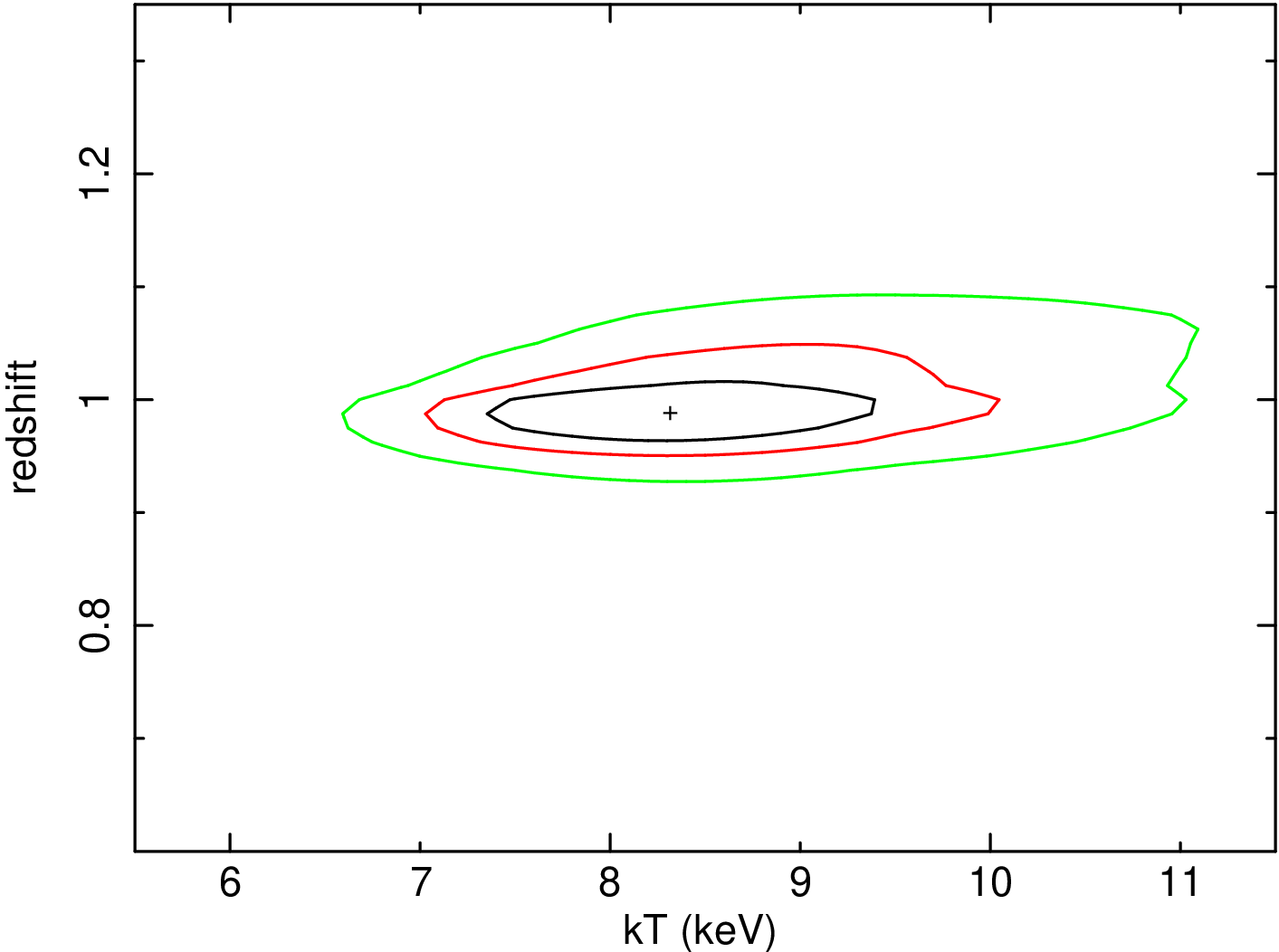}
   \hspace{0.05\textwidth}
   \includegraphics[width=0.33\textwidth,angle=-90]{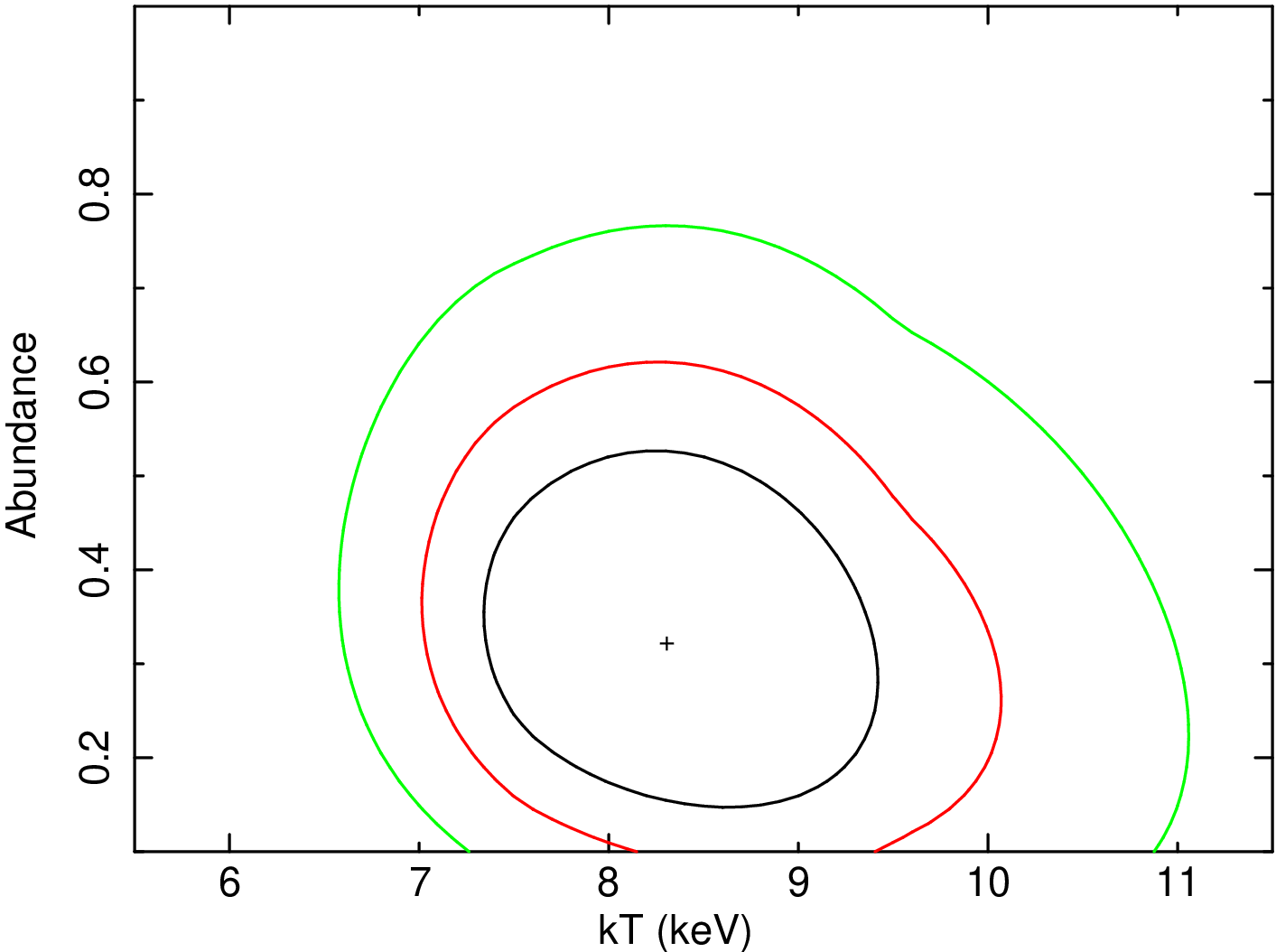}
   \end{center}
   \caption{
      {\em Left}: 
      Confidence contours for redshift and temperature determined from
      the X-ray spectra. Starting from the innermost contour, the levels
      indicate 68\,\%, 90\,\%, and 99\,\% confidence. 
      {\em Right:}
      Same confidence contours for metal abundance and temperature.
      } 
   \label{contours} 
   \end{figure*}

   We used \xspec version 12.0 to subtract the background spectra and to fit  \mekal plasma models.
   The  \mekal code \citep{Kaastra,Liedahl} allows to set 
   a metal abundance parameter relative to the solar values provided by \citet{Anders}.
   Fits were carried out using the
   Cash-statistics and a binning with at least one photon per spectral bin. The
   Galactic neutral hydrogen column density in the direction of \clshort
   is $N_{\rm H} = 4.0 \times 10^{20} \: {\rm cm^{-2} }$, as obtained
   from radio data \citep{dickey:90}. The Fe\,K-line is visible in both
   spectra at $k_{\rm B}T \sim 3 \, {\rm keV}$, see \Fig{spectrum}.
   Hence, the redshift of \clshort is unambiguously determined by the
   X-ray data, see \Fig{contours}. The best-fit redshift, temperature,
   and metallicity are $z=0.99\pm0.03$, $k_{\rm B}T=8.2\pm0.9\,{\rm
   keV}$, and  $Z/Z_\odot = 0.32 \pm 0.18 $. The latter is in agreement with
   the results of  \citet{balestra:07} and \citet{maughan:08}, who analysed the 
   {\em Chandra} spectra of larger cluster samples and found
   $<Z/Z_\odot> = 0.28$ and  $<Z/Z_\odot> = 0.2$ as the mean abundances at $z\sim 1$.

   \begin{figure}
   \begin{center}
   \includegraphics[width=0.45\textwidth,clip]{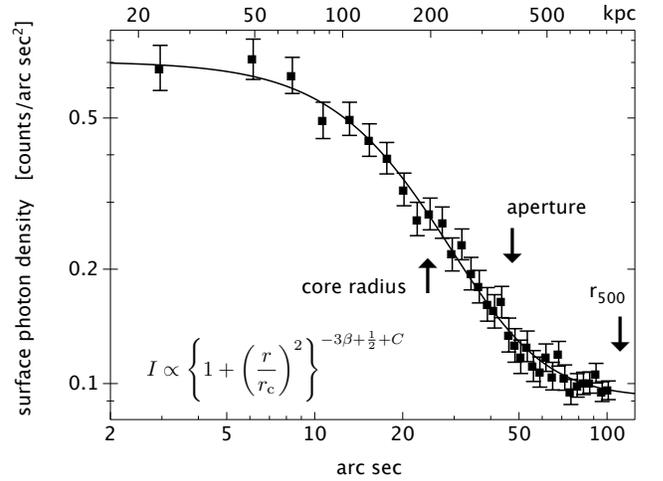}
   \end{center}
   \caption{
      Spherically averaged surface luminosity profile used to
      extrapolate the X-ray luminosity to $r_{500}$.
      } 
   \label{beta} 
   \end{figure}

    The spectral fit provides also the background subtracted bolometric
   X-ray flux $F_{\rm bol}=2.4 \times 10^{-13} \: {\rm erg \, s^{-1} \,
   cm^{-2}}$ within the aperture of $45''$. Using the concordance
   cosmological model allows to calculate the intrinsic luminosity of
   the cluster. However, the used aperture corresponds to a radius
   significantly smaller than $r_{500}$. We estimate the
   missing flux by approximating a $\beta$-model. \Fig{beta} shows the
   spherically averaged surface photon density for the MOS1 image.
   Allowing all parameters to vary in a least-square fit we find  for the
   slope $\beta = 0.73 \pm 0.1$ and for the core radius $r_{\rm c} =
   (24 \pm 4) ''$. Now, we determine $r_{500}$ by the temperature-radius
   relation given in \citet{ohara:07}. For the redshift and temperature
   of \clshort derived above we obtain $r_{500} = (860\,\pm 50) \,{\rm kpc}$,
   corresponding to $(107 \pm 6) ''$. With the results of the
   $\beta$-profile fit we find for the intrinsic luminosity of \clshort
   $L_{\rm bol}(<r_{500}) = 1.8 \times 10^{45} \: {\rm erg \, s^{-1} } $.
   \\

   The mass of a cluster can be estimated by a model based on spherical
   symmetry and hydrostatic equilibrium. Since for \clshort the
   determination of the temperature profile is not possible, we
   additionally assume isothermality. As a result, the total mass is
   given by
   \begin{equation}
     M_{500}
     \sim
     \frac{3 \beta}{G} \,
     \frac{ k_{\rm B} T \, r_{500} }  { \mu m_{\rm p} }
     \frac{(r/r_{\rm c})^2}{1+(r/r_{\rm c})^2}
   \end{equation}

   \citep[see e.g][]{hicks:07}.

   Using the properties of \clshort as derived above we obtain for the mass
   $M_{500} = ( 5.6 \pm 1.0  ) \times 10^{14} \: M_\odot$, where we set
   $\mu=0.6$ as expected for a fully ionised ICM. For comparison, the
   mass-temperature relation of \citet{vikhlinin:06} predicts 
   $(3.8 \pm 0.7 ) \times 10^{14} \: M_\odot$, which is within $2\sigma$ of
   our derived value.
   \\
   
   The luminosity of this cluster agrees within the errors with
   the local $L-T$ relation \citep{markevitch:98}. However, \cite{kotov:05} and
   \cite{maughan:06} argued that the luminosity scales with redshift
   roughly in concordance with the expectations of self-similar cluster
   evolution. Contrarily,  \cite{ohara:07} found only very
   little evolution of the $L_{500}-T$-relation in a recent analysis of a sample of 70
   clusters observed with {\em Chandra}. A similar
   discrepancy has been found for the most distant cluster, 
   \object{XMMXCS J2215.9-1738}; it is under-luminous even compared to the local
   $L_{500}-T$-relation \citep{hilton:07}. This illustrates that the 
   evolution of gas scaling properties is still poorly constrained. Hence, a larger 
   high redshift cluster sample is needed and selection effects have to
   be modelled properly.

\section{Optical imaging}
\label{lbt}

\begin{figure}
\includegraphics[width=8.8cm]{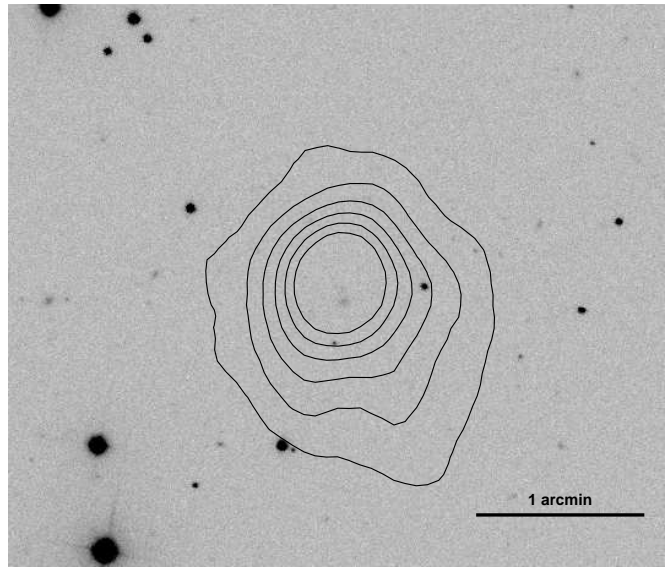}
\caption{SDSS i-band image with EPIC MOS1 X-ray contours of \clshort overlaid. The object 6 arcsec south
south of the X-ray position is \object{SDSS J083026.39+524130.5}.} 
\label{sdss} 
\end{figure}

No cluster galaxies or any other plausible optical counterparts for an extended X-ray
source are visible in SDSS images at the position of \clshort (Fig. \ref{sdss}).  A rather blue galaxy 
(\object{SDSS J083026.39+524130.5}, $r-z=0.63$) with disk-like 
appearance  some 6 arcsec south of the X-ray position is most likely a foreground spiral galaxy.

\begin{figure}
\centerline{\includegraphics[width=8.8cm]{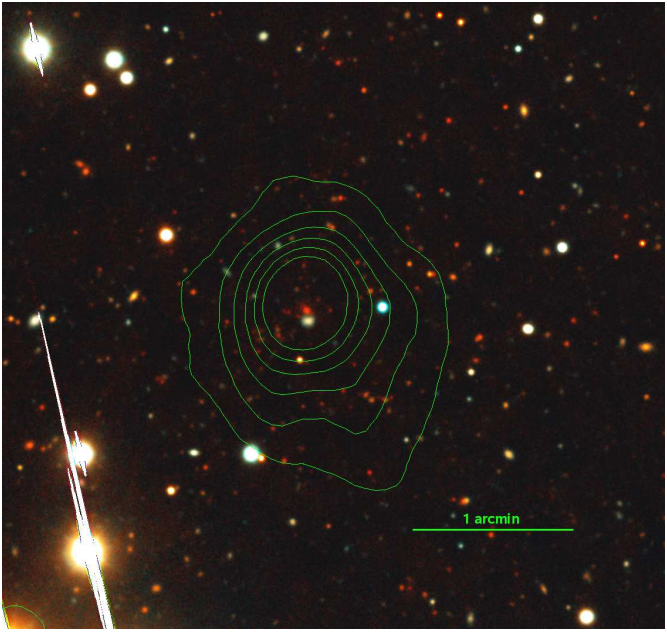}}
\caption{LBT $r$- and $z$-band colour composite image with XMM EPIC contours (green). The blue galaxy just south
of the X-ray centre is probably unrelated to the cluster and corresponds to  \object{SDSS J083026.39+524130.5}.} 
\label{lbtimage} 
\end{figure}

We have obtained deep  $r_{\rm sloan}$ and $z_{\rm sloan}$ images in the field of \clshort using the LBC cameras 
of the LBT. The $r$-band image was taken with the blue LBC camera and the $z$-band image with
the red camera. Both bands have total exposure times of 18 minutes ($9 \times 2$ minutes) 
and were taken on May 10th, 2008. The data reduction was accomplished  with a pipeline software based on
the Garching-Bonn Deep Survey (GaBoDS) pipeline, which has been modified
at AIP to process the LBC data.
The astrometry and photometry of the stacked images has been tied to the SDSS.
The image quality of the reduced images is FWHM=1.1 arcsec in the $z$-band image and 1.5 arcsec in the $r$-band.
A colour composite of the two images is shown in Fig. \ref{lbtimage}.  
The LBT images reveal a clear overdensity of red galaxies within the X-ray contours.
The brightest cluster galaxy is located close to the peak of the X-ray emission.
Its z-band magnitude of 19.44 and colour $r-z$=2.15 are consistent
with the X-ray redshift $z=1$.
A detailed analysis of the LBT data will be presented in a forthcoming paper.

\section{Summary}
\label{conclusion}

Based on the X-ray data we show that \clxmm is a
massive cluster at $z \sim 1$.
Its luminosity of  $1.8 \times 10^{45} {\rm erg\,  s^{-1}}$ and temperature of  $8.2 \pm 0.8$ keV are unsurpassed 
at its distance and beyond.
The only comparable known cluster is \clrosat \citep{ebeling} which is less distant at $z=0.89$,
but even more luminous and hotter 
\citep[$L_{\rm bol} = 5 \times 10^{45} {\rm erg \, s^{-1}}$, ${\rm kT = 10.4\; keV}$,][]{maughan07}.

The luminosity of  \clshort is possibly enhanced by a cool core.
Since its X-ray flux makes \clshort an ideal target for more detailed 
investigations in X-rays, we have applied for a deep {\em Chandra} observation in order to 
investigate a possible  cool core of the cluster and to establish an improved mass model.

How likely is it to find a cluster with the mass of \clshort in the  \xmm serendipitous survey data?
We have estimated the number of clusters with $M \geq 5.6 \times 10^{14} M_{\odot}$ 
expected in the redshift shell $0.9<z<1.1$ and $0.3\%$ of the sky corresponding to our 
survey area in the 2XMM-SDSS overlap.
The mass function of \citet{gottloeber}, based on the large {\em MareNostrum Universe} SPH
simulation using the ``{\em WMAP} first year'' cosmology ($\sigma_8=0.9$) predicts 5 clusters.
Calculating the mass function following \citet{PS} for the ``{\em WMAP} 3 years'' 
cosmological parameters with $\sigma_8=0.76$ predicts 0.1 clusters in our survey.
Changing $\sigma_8$ to $0.82$ increases the expected cluster count to 0.4.
Given the uncertainty of our mass estimate, we conclude  that the \xmm discovery of one or few
clusters like \clshort  is within the expectations. 
However, we will have to await the  {\em eROSITA } all-sky survey \citep{predehl} to 
find a larger number of these very massive clusters.

\begin{acknowledgements}
    GL and MH acknowledge support by the Deutsches Zentrum f\"ur Luft- und Raumfahrt
    (DLR) under contracts 50 QR 0802 and 50 OX 0201. JK is supported by the DFG priority programme
    SPP1177 (grant no.~SCHW563/23-1). This work is based on observations obtained with XMM-Newton 
    and on data acquired using the Large Binocular Telescope (LBT). We thank Alexander Knebe for providing
    code to calculate the Press-Schechter mass function.\\
\end{acknowledgements}

\end{document}